\documentclass[preprint,showpacs,preprintnumbers,amsmath,amssymb,superscriptaddress]{revtex4}

\usepackage{graphicx}
\usepackage{dcolumn}
\usepackage{bm}

\font\scripti=cmmi7
\font\scriptscripti=cmmi5
\def\sib#1{\setbox0 = \hbox{\scripti #1}
  \kern-.02em\copy0\kern-\wd0
  \kern.04em\box0} 
\def\ssib#1{\setbox0 = \hbox{\scriptscripti #1}
  \kern-.02em\copy0\kern-\wd0
  \kern.04em\box0} 
\font\tenib=cmmib10 
\skewchar\tenib='177 \skewchar\tenib='177 \skewchar\tenib='177
\def\pbold#1{\setbox0 = \hbox{$ #1 $}
  \kern-.022em\copy0\kern-\wd0
  \kern.011em\copy0\kern-\wd0
  \kern.011em\copy0\kern-\wd0
  \kern.011em\copy0\kern-\wd0
  \kern.011em\box0} 

\usepackage{graphicx}
\usepackage{dcolumn}
\usepackage{bm}

\def\lesssim{\ \raise.3ex\hbox{$<$}\kern-0.8em\lower.7ex\hbox{$\sim$}\ }
\def\gesim{\ \raise.3ex\hbox{$>$}\kern-0.8em\lower.7ex\hbox{$\sim$}\ }

\def\dag{\dagger}

\def\sig{\sigma}
\def\Sig{\Sigma}
\def\om{\omega}
\def\Gam{\Gamma}

\begin{document}
\title{Hetero pairing and component-dependent pseudogap phenomena in an ultracold Fermi gas with mass imbalance}
\author{Ryo Hanai and Yoji Ohashi}
\affiliation{Department of Physics, Keio University, 3-14-1 Hiyoshi, Kohoku-ku, Yokohama 223-8522, Japan} 
\date{\today}
\begin{abstract}
We investigate the superfluid phase transition and single-particle excitations in the BCS (Bareen-Cooper-Schrieffer)-BEC (Bose-Einstein condensation) crossover regime of an ultracold Fermi gas with mass imbalance. In our recent paper [R. Hanai, \textit{et. al.}, Phys. Rev. A {\bf 88}, 053621 (2013)], we showed that an extended $T$-matrix approximation (ETMA) can overcome the serious problem known in the ordinary (non-self-consistent) $T$-matrix approximation that it unphysically gives double-valued superfluid phase transition temperature $T_{\rm c}$ in the presence of mass imbalance. However, at the same time, the ETMA was also found to give the vanishing $T_{\rm c}$ in the weak-coupling and highly mass-imbalanced case. In this paper, we inspect the correctness of this ETMA result, using the self-consistent $T$-matrix approximation (SCTMA). We show that the vanishing $T_{\rm c}$ is an artifact of the ETMA, coming from an internal inconsistency of this theory. The superfluid phase transition actually always occurs, irrespective of the ratio of mass imbalance. We also apply the SCTMA to the pseudogap problem in a mass-imbalanced Fermi gas. We show that pairing fluctuations induce different pseudogap phenomena between the the light component and heavy component. We also point out that a $^6$Li-$^{40}$K mixture is a useful system for the realization of a hetero pairing state, as well as for the study of component-dependent pseudogap phenomena.
\end{abstract}
\pacs{03.75.Ss, 03.75.-b, 67.85.-d}
\maketitle
\section{Introduction}\label{sec1}
\par
The realization of an unconventional superfluid state beyond the $^{40}$K\cite{Regal2004} and $^6$Li\cite{Zwierlein2004,Kinast2004,Bartenstein2004} superfluid Fermi gases is one of the most exciting challenges in cold Fermi gas physics. Although no one has succeeded in this attempt, various possibilities have been so far explored, such as a $p$-wave superfluid\cite{Regal2003,Zhang2004,Schunck2005,Ohashi2005,Ho2005,Gurarie2005,Levinsen2007,Iskin2005,Inotani2012}, the Berezinskii-Kosterlitz-Thouless state\cite{Feld2011,Sommer2012,Iskin2009,Watanabe2013,Bauer2014}, a superfluid state with hetero-Cooper-pairs\cite{Liu2003,Forbes2005,Wille2008a,Taglieber2008,Voigt2009,Costa2010,Naik2010,Spiegelhalder2010,Tiecke2010,Lin2006,Wu2006,Iskin2006,Iskin2007,Pao2007,Parish2007,Orso2008,Baranov2008,Gezerlis2009,Diener2010,Stoof1,Stoof2,Takemori2012,Lan2013,Hanai2013}, the Sarma phase\cite{Stoof1,Stoof2,Sarma1963,Sheehy2007}, a Fermi superfluid with a spin-orbit interaction\cite{Lin2011,Wang2012,Cheuk2012,Jiang2011}, and a dipolar Fermi superfluid\cite{Baranov,Endo}. Once one of them is realized, one could clarify its  superfluid properties, maximally using the high tunability of Fermi gases\cite{Chin2010} and various experimental techniques\cite{Chin2004,Stewart2008,Gaebler2010,Ketterle2008,Salomon2010,Ketterle2011,Martin2012}. Since an ultracold Fermi gas is expected as a useful quantum simulator for strongly interacting Fermi systems, this challenge would also be important on the viewpoint of this application. 
\par
Among various possibilities, we pick up the hetero-pairing state\cite{Liu2003,Forbes2005,Wille2008a,Taglieber2008,Voigt2009,Costa2010,Naik2010,Spiegelhalder2010,Tiecke2010,Lin2006,Wu2006,Iskin2006,Iskin2007,Pao2007,Parish2007,Orso2008,Baranov2008,Gezerlis2009,Diener2010,Stoof1,Stoof2,Takemori2012,Lan2013,Hanai2013} in this paper. This unconventional superfluid state is expected in a $^6$Li-$^{40}$K mixture, and is characterized by Cooper pairs composed of different species\cite{Liu2003,Forbes2005,Wille2008a,Taglieber2008,Voigt2009,Costa2010,Naik2010,Spiegelhalder2010,Tiecke2010,Lin2006,Wu2006,Iskin2006,Iskin2007,Pao2007,Parish2007,Orso2008,Baranov2008,Gezerlis2009,Diener2010,Stoof1,Stoof2,Takemori2012,Lan2013,Hanai2013}. Although the superfluid phase transition of this Fermi-Fermi mixture has not been reported yet, the Fermi degenerate regime has been achieved\cite{Taglieber2008,Spiegelhalder2010}. In addition, since a tunable interaction associated with a Feshbach resonance between $^6$Li and $^{40}$K atoms\cite{Wille2008a,Costa2010,Naik2010}, as well as the formation of $^6$Li-$^{40}$K hetero molecules\cite{Voigt2009}, has been realized, the observation of superfluid behaviors seems imminent. Since the condensation of hetero pairs is also discussed in, for example, an exciton gas\cite{Yoshioka2011,Stolz2012,Yoshioka2013,High2012,Versteegh2012}, an exciton-polariton gas\cite{Imamoglu1996,Tassone1999,Deng2002,Kasprzak2006}, as well as a dense quark matter\cite{Barrois1977,Bailin1984}, the realization of a superfluid $^6$Li-$^{40}$K Fermi gas would give great impact on these fields. 
\par
\begin{figure}
\begin{center}
\includegraphics[width=0.5\linewidth,keepaspectratio]{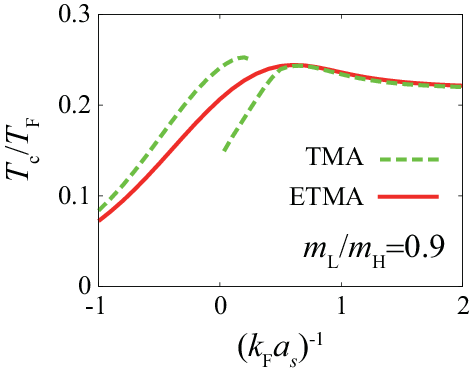}
\end{center}
\caption{(Color online) Calculated superfluid phase transition temperature $T_{\rm c}$ in a mass-imbalanced Fermi gas\cite{Hanai2013}. We take $m_{\rm L}/m_{\rm H}=0.9$, where $m_{\rm L}$ ($m_{\rm H}$) is a mass of a light (heavy) atom. TMA: (non-self-consistent) $T$-matrix approximation. ETMA: Extended $T$-matrix approximation. As usual, the interaction strength is measured in terms of the inverse scattering length $(k_{\rm F}a_s)^{-1}$ (where $k_{\rm F}$ is the Fermi momentum). The temperature is normalized by the Fermi temperature $T_{\rm F}=k_{\rm F}^2/(2m)$, where $m^{-1}=[m_{\rm L}^{-1}+m_{\rm H}^{-1}]/2$.}
\label{fig1}
\end{figure}
\par
\begin{figure}
\begin{center}
\includegraphics[width=0.9\linewidth,keepaspectratio]{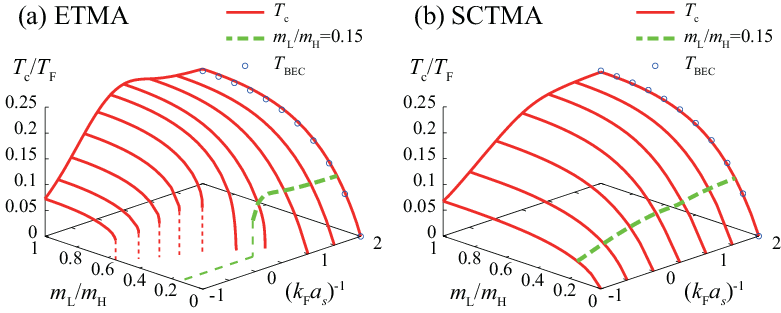}
\end{center}
\caption{(Color online) Calculated $T_{\rm c}$ as functions of the interaction strength $(k_{\rm F}a_{\rm s})^{-1}$ and the ratio $m_{\rm L}/m_{\rm H}$ of mass imbalance. (a) Extended $T$-matrix approximation (ETMA). (b) Self-consistent $T$-matrix approximation (SCTMA). The dashed line shows the case of a $^6$Li-$^{40}$K mixture ($m_{\rm L}/m_{\rm H}=6/40=0.15$). The open circles are the BEC phase transition temperature $T_{\rm BEC}$ in an ideal molecular Bose gas, given by Eq. (\ref{BEC}).}
\label{fig2}
\end{figure}

In the current stage of research for the hetero Fermi superfluid, the evaluation of the superfluid phase transition temperature $T_{\rm c}$ is a crucial theoretical issue. In our recent paper\cite{Hanai2013}, we showed that the ordinary (non-self-consistent) $T$-matrix approximation (TMA)\cite{Perali2002}, which has been extensively used to successfully clarify various interesting BCS-BEC crossover physics in the mass-{\it balanced} case\cite{Stewart2008,Gaebler2010,Chen2009,Tsuchiya2009,Tsuchiya2010,Watanabe2010,Hu2010,Watanabe2012}, breaks down in the presence of mass imbalance, to unphysically give double-valued $T_{\rm c}$ in the crossover region, as shown in Fig. \ref{fig1}. In Ref. \cite{Hanai2013}, we overcame this difficulty by employing an extended $T$-matrix approximation (ETMA)\cite{Kashimura2012,Tajima2014}, which involves higher order pairing fluctuations beyond the TMA. However, apart from the recovery of the expected single-valued $T_{\rm c}$ (see Fig \ref{fig1}), the ETMA was found to give vanishing $T_{\rm c}$ in the BCS regime when $m_{\rm L}/m_{\rm H}\ll 1$, as shown in Fig. \ref{fig2}(a) (where $m_{\rm L}$ ($m_{\rm H}$) is a mass of a light (heavy) atom). Since this predicts that a $^6$Li-$^{40}$K mixture has a critical interaction strength, below which the superfluid instability is absent (see the dotted line in Fig. \ref{fig2}(a)), it is a crucial issue to inspect the correctness of this.
\par
\par
In this paper, we extend the ETMA to the self-consistent $T$-matrix approximation (SCTMA)\cite{Haussmann1999,Zwerger,Enss}, to calculate $T_{\rm c}$ in a mass-imbalanced Fermi gas. We clarify that the vanishing $T_{\rm c}$ seen in Fig. \ref{fig2}(a) is an artifact, originating from an internal inconsistency of the ETMA. As shown in Fig. \ref{fig2}(b), the superfluid phase transition actually always occurs in the presence of mass imbalance, which is one of our main results in this paper.
\par
Using the SCTMA, we also examine single-particle properties of a mass-imbalanced Fermi gas. As in the mass-balanced case, this system is found to exhibit the pseudogap phenomenon in the BCS-BEC crossover region. However, details of this many-body phenomenon are shown to be different between light atoms and heavy atoms. Since such a component-dependent pseudogap phenomenon never occurs in a mass-balanced Fermi gas, it is characteristic of a Fermi gas with mass imbalance. 
\par
This paper is organized as follows. In Sec. II, we explain the self-consistent $T$-matrix approximation in the presence of mass imbalance. In Sec. III, we examine $T_{\rm c}$. Here, we explain why the ETMA incorrectly gives the vanishing $T_{\rm c}$ in the highly mass-imbalanced regime, as well as the reason why this problem is solved in the SCTMA. In Sec. \ref{sec4}, we calculate the single-particle density of states, as well as the single-particle spectral weight, to see how pseudogap phenomena differently appear in the light component and heavy component. In Sec. IV, we consider the case of a $^6$Li-$^{40}$K mixture. Throughout this paper, we set $\hbar=k_{\rm B}=1$, and the system volume $V$ is taken to be unity, for simplicity.
\par
\begin{figure}
\begin{center}
\includegraphics[width=0.65\linewidth,keepaspectratio]{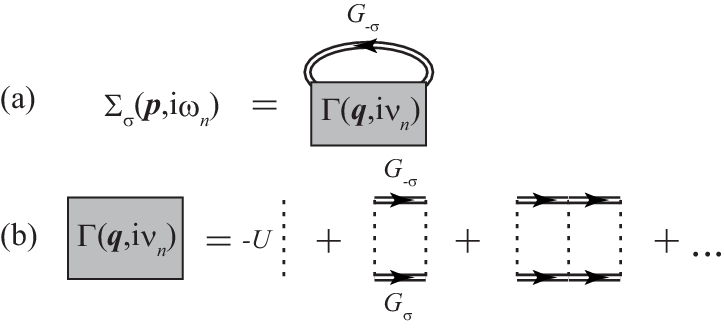}
\end{center}
\caption{(a) Self-energy $\Sigma_\sigma(\bm p,i\omega_n)$ in the self-consistent $T$-matrix approximation (SCTMA). The double solid line is the dressed Green's function $G_\sigma$ in Eq. (\ref{sctmag}). (b) Particle-particle scattering matrix $\Gamma(\bm q,i\nu_n)$ is the SCTMA. The dotted line describes the pairing interaction $-U$. In this figure, $-\sigma$ means the opposite component to $\sigma={\rm L},{\rm H}$.
}
\label{fig3}
\end{figure}
\par\par
\section{Formulation}\label{sec2}
\par
We consider a two-component Fermi gas with mass imbalance, described by the BCS-type Hamiltonian,
\begin{equation}
H = \sum_{\bm{p},\sigma={\rm L,H}}\xi_{\bm{p},\sig}c^{\dag}_{\bm p,\sig}c_{\bm p,\sig}
-U\sum_{\bm q}\sum_{\bm p, \bm{p}'}
c^{\dag}_{\bm p + \bm q/2,{\rm L}}c^\dag_{-\bm p + \bm q/2,{\rm H}}c_{-\bm {p}' + \bm q /2,{\rm H}}c_{\bm{p}'+\bm q /2,{\rm L}}.
\label{HAM}
\end{equation}
Here, $c_{{\bm p},{\rm L}}$ and $c_{{\bm p},{\rm H}}$ describe a light atom with a mass $m_{\rm L}$ and a heavy atom with a mass $m_{\rm H}$, respectively. $\xi_{\bm p, \sigma}=p^2/(2m_\sigma)-\mu_{\sigma}$ ($\sigma={\rm L,H}$) is the kinetic energy of a Fermi atom, measured from the Fermi chemical potential $\mu_\sigma$. $-U~(<0)$ is a pairing interaction, which is related to the $s$-wave scattering length $a_s$ as,
\begin{equation}
\frac{4\pi a_s}{m}=\frac{-U}{1-U\sum_{\bm p}\frac{m}{p^2}},
\label{eq.as}
\end{equation}
where $m=2m_{\rm L}m_{\rm H}/(m_{\rm L}+m_{\rm H})$ is twice the reduced mass. As in the mass-balanced case, we measure the interaction strength in terms of $a_s$ in this paper. The weak-coupling BCS regime and the strong-coupling BEC regime are then characterized by $(k_{\rm F}a_s)^{-1}\lesssim -1$ and  $1 \lesssim (k_{\rm F}a_s)^{-1}$, respectively (where $k_{\rm F}=(3\pi^2 N)^{1/3}$ is the Fermi momentum, and $N$ is the total number of Fermi atoms). The BCS-BEC crossover region is given by $-1 \lesssim (k_{\rm F}a_s)^{-1}\lesssim 1$.
\par
In this paper, we measure the momentum $p$, energy $\omega$, and temperature $T$, in terms of, respectively, the Fermi momentum $k_{\rm F}=(3\pi^2 N)^{1/3}$, Fermi energy $\varepsilon_{\rm F}=k_{\rm F}^2/(2m)$, and Fermi temperature $T_{\rm F}=\varepsilon_{\rm F}$, of a mass-balanced free Fermi gas with the atomic mass $m=2m_{\rm L}m_{\rm H}/(m_{\rm L}+m_{\rm H})$ and the particle number $N$. We briefly note that, while $k_{\rm F}$ remains unchanged in a mass-imbalanced Fermi gas, $\varepsilon_{\rm F}$ is different from the Fermi energy $\varepsilon_{\rm F}^\sigma=k_{\rm F}^2/(2m_\sigma)$ of each component in the presence of mass imbalance. 
\par
The single-particle thermal Green's function is given by,  
\begin{equation}
G_\sigma({\bm p},i\omega_n)=\frac{1}{i\om_n - \xi _{\bm p,\sig}-\Sig_\sig(\bm p,i\om_n)},
\label{sctmag}
\end{equation}
where $\omega_n$ is the fermion Matsubara frequency. The self-energy $\Sig_\sig(\bm p,i\om_n)$ describes strong-coupling corrections to single-particle excitations. In the SCTMA\cite{Haussmann1999}, $\Sig_\sig(\bm p,i\om_n)$ is diagrammatically described as Fig. \ref{fig3}. Summing up the diagrams, we obtain
\begin{equation}
\Sigma_\sigma({\bm p},i\om _n)=T\sum _{\bm q,\nu _n}\Gam(\bm q,i\nu _n)
G_{-\sig}({\bm q}-{\bm p},i\nu_n-i\om_n).
\label{sctma}
\end{equation}
Here, $\nu_n$ is the boson Matsubara frequency, and $-\sigma$ denotes the opposite component to $\sigma={\rm L},{\rm H}$. $\Gamma(\bm q,i\nu_n)$ is the particle-particle scattering matrix describing fluctuations in the Cooper-channel, which is given by, in the SCTMA,
\begin{equation}
\Gam(\bm q, i\nu _n)={-U  \over 1-U\Pi(\bm q,i\nu_n)},
\label{GAM}
\end{equation}
where
\begin{eqnarray}
\Pi(\bm q, i\nu_n)&=&T\sum_{\bm p,i\om _n}G_{\rm L}(\bm p + \bm q/ 2,i\nu_n+i\om_n)G_{\rm H}(-\bm p+\bm q /2,-i\om_n)
\label{PI}
\end{eqnarray}
is the pair correlation function.
\par
As usual, we determine $T_{\rm c}$ from the Thouless criterion $\big[\Gamma (\bm q=0,i\nu_n=0) \big]^{-1}=0$\cite{Haussmann1999,Thouless1960}, which gives
\begin{equation}
1=U\Pi(\bm q=0,i\nu_n=0).
\label{GAP}
\end{equation}
We solve this $T_{\rm c}$ equation, together with the equations for the number $N_\sigma=N/2$ of Fermi atoms in the $\sigma$ component,
\begin{equation}
N_\sigma =T\sum_{\bm p,i\omega_n}G_{\bm p\sigma}(i\omega_n),
\label{number}
\end{equation}
to self-consistently determine $(T_{\rm c},\mu_{\rm L},\mu_{\rm H})$. Above $T_{\rm c}$, we only solve the number equation (\ref{number}), to determine $(\mu_{\rm L}, \mu_{\rm H})$.
\par
We note that the SCTMA is a consistent theory in the sense that the dressed Green's function $G_\sigma$ in Eq. (\ref{sctmag}) is used everywhere in the diagrams in Fig. \ref{fig3}. In this sense, the ETMA employed in our previous paper\cite{Hanai2013} has an internal inconsistency. That is, while the dressed Green's function is used in the fermion loop in Fig. \ref{fig3}(a), the bare Green's function, 
\begin{equation}
G_\sigma^0(\bm p,i\omega_n)={1 \over i\omega_n-\xi_{\bm p \sigma}},
\label{GAM0}
\end{equation}
is used in the particle-particle scattering matrix $\Gamma({\bm q},i\nu_n)$ in Fig. \ref{fig3}(b). Because of this, the ETMA pair correlation function $\Pi({\bm q},i\nu_n)$ in Eq. (\ref{GAM}) is in the lowest order with respect to the pairing interaction $-U$ as 
\begin{eqnarray}
\Pi_{\rm ETMA}(\bm q, i\nu_n)&=&T\sum_{\bm p,i\om _n}G^0_{\rm L}(\bm p + \bm q/ 2,i\nu_n+i\om_n)G^0_{\rm H}(-\bm p+\bm q /2,-i\om_n)
\nonumber\\ 
&=&-\sum_{\bm p}\frac{1-f(\xi_{\bm{p}+\bm{q}/2,{\rm L}})-f(\xi_{-\bm{p}+\bm{q}/2,{\rm H}})}{i\nu_n-\xi_{\bm{p}+\bm{q}/2,{\rm L}}-\xi_{-\bm{p}+\bm{q}/2,{\rm H}}}.
\label{PIetma}
\end{eqnarray}
Here, $f(\varepsilon)=[e^{\varepsilon/T}+1]^{-1}$ is the Fermi distribution function. Thus, although the number equations in the ETMA use the dressed Green's function involving strong-coupling corrections, the Thouless criterion, $[\Gamma_{\rm ETMA}(0, 0)]^{-1}=0$, gives the BCS-type $T_{\rm c}$-equation,
\begin{equation}
1={U \over 2}\sum_{\bm p}
{\displaystyle \tanh{\xi_{\bm{p},{\rm L}} \over 2T}+\tanh{\xi_{\bm{p},{\rm H}} \over 2T}
\over
\xi_{\bm{p},{\rm L}}+\xi_{\bm{p},{\rm H}}
},
\label{gapETMA}
\end{equation}
where $\Gamma_{\rm ETMA}(\bm{q},i\nu_n)=(-U)/(1-U\Pi_{\rm ETMA}(\bm{q},i\nu_n))$. In Sec. III, we will find that this inconsistent treatment is the origin of the vanishing $T_{\rm c}$ seen in Fig. \ref{fig2}(a). We briefly note that, when we replace all the dressed Green's functions in Fig. \ref{fig3} by the bare ones, the ordinary non-self-consistent $T$-matrix approximation\cite{Perali2002} is recovered.
\par
We also examine strong-coupling corrections to single-particle excitations in a mass-imbalanced Fermi gas. As usual, we calculate the single-particle density of states $\rho_\sigma(\omega)$, as well as the single-particle spectral weight $A_\sigma({\bm p},\omega)$, from the SCTMA Green's function in Eq.  (\ref{GAM}) as,
\begin{equation}
\rho_\sigma(\omega)=-{1 \over \pi} \sum_{\bm p} {\rm Im} \big[G_\sigma(\bm p,i\omega_n \rightarrow \omega+i\delta) \big],
\label{DOS}
\end{equation}
\begin{equation}
A_\sigma({\bm p},\omega)=-{1 \over \pi}
{\rm Im}
[G_\sigma({\bm p},i\omega_n\to \omega+i\delta],
\label{sp}
\end{equation}
where $\delta$ is an infinitesimally small positive number. The density of states $\rho_\sigma(\omega)$ equals the momentum summation of the spectral weight $A_\sigma({\bm p},\omega)$ for a given energy $\omega$.
\par
\begin{figure}
\begin{center}
\includegraphics[width=0.45\linewidth,keepaspectratio]{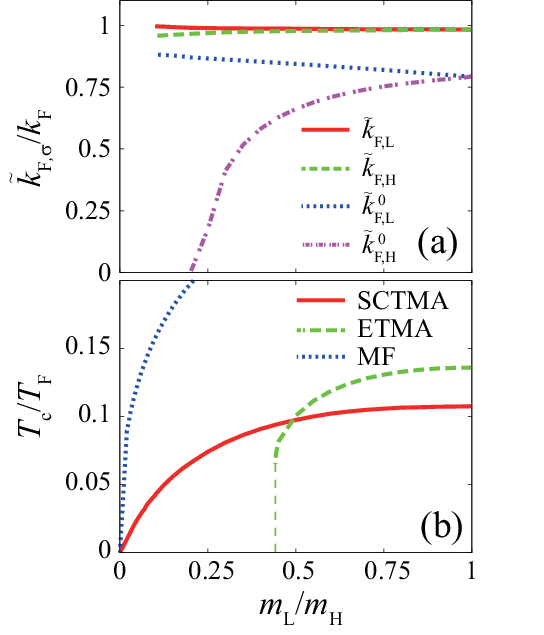}
\end{center}
\caption{(Color online) (a) Effective Fermi momenta ${\tilde k}_{{\rm F}.\sigma}$ in the SCTMA. We take $(k_{\rm F}a_s)^{-1}=-0.5$. We also show ${\tilde k}_{{\rm F},\sigma}^0=\sqrt{2m_\sigma\mu_\sigma}$ in the SCTMA. The ETMA also gives almost the same result for ${\tilde k}_{{\rm F},\sigma}^0$, although we do not explicitly show it here. (b) $T_{\rm c}$ when $(k_{\rm F}a_s)^{-1}=-0.5$. SCTMA: Self-consistent $T$-matrix approximation. ETMA: Extended $T$-matrix approximation. MF: Mean-field approximation.
}
\label{fig4}
\end{figure}
\par
\par
\section{Superfluid phase transition and effects of mass imbalance}\label{sec3}
\par
As we have already shown in Fig. \ref{fig2}(b), the SCTMA always gives a finite $T_{\rm c}$, even in the presence of mass imbalance. Thus, the BCS-BEC crossover phenomenon, which has been already observed in $^{6}$Li and $^{40}$K Fermi gases, is also expected in a $^6$Li-$^{40}$K mixture ($m_{\rm L}/m_{\rm H}=0.15$). We emphasize that the ETMA gives the different prediction that this Fermi-Fermi mixture does not exhibit the superfluid phase transition in the BCS regime\cite{Hanai2013}.
\par
To explain the reason for this difference, we introduce the effective Fermi momentum ${\tilde {\bm k}}_{{\rm F},\sigma}$, which is determined from the equation for the pole of the analytic continued dressed Green's function $G_\sigma({\bm p},i\omega_n\to \omega+i\delta)$ at $\omega=0$,
\begin{equation}
{{\tilde {\bm k}}_{{\rm F},\sigma}^2 \over 2m_\sigma}-\mu_\sigma
+ {\rm Re}\big{[} \Sigma_\sigma({\tilde {\bm k}}_{{\rm F},\sigma},i\omega_n
\to \omega+i\delta=0+i\delta)\big{]}=0.
\label{kFstar}
\end{equation}
For a free Fermi gas at $T=0$, ${\tilde k}_{{\rm F},\sigma}$ just equals the Fermi momentum $k_{\rm F}=(3\pi^2N)^{1/3}$. Thus, this quantity physically describes the size of a Fermi surface in the $\sigma$ component\cite{note}. As shown in Fig. \ref{fig4}(a), the SCTMA gives ${\tilde k}_{{\rm F},{\rm L}}\simeq{\tilde k}_{{\rm F},{\rm H}}$, indicating that the Fermi surface in the light component has almost the same size as that in the heavy component, irrespective of the ratio $m_{\rm L}/m_{\rm H}$. In a sense, this is reasonable, because the number $N/2$ of Fermi atoms in the $\sigma$ component is roughly estimated as  $N/2\sim (4\pi {\tilde k}_{{\rm F},\sigma}^3/3)/(2\pi/L)^3$ (where $L$ is the system size), which is independent of the atomic mass $m_\sigma$. Since the superfluid phase transition in the BCS regime is dominated by the pair formation between a light atom with the momentum ${\bm p}~(\simeq{\tilde {\bm k}}_{{\rm F},{\rm L}})$ and a heavy atom with $-{\bm p}~(\simeq -{\tilde {\bm k}}_{{\rm F},{\rm H}})$, the (approximate) coincidence of two Fermi surfaces is favorable to the superfluid instability. As a result, the SCTMA, which consistently uses the dressed Green's function $G_\sigma({\bm p},i\omega_n)$ in both the $T_{\rm c}$-equation (\ref{GAP}) and the number equation (\ref{number}), always gives a finite $T_{\rm c}$, as shown in Fig. \ref{fig4}(b). 
\par
On the other hand, the ETMA uses the bare Green's function $G^0_\sigma({\bm p},i\omega_n)$ in the $T_{\rm c}$-equation (\ref{gapETMA}). Thus, while the coincidence of the two Fermi surfaces is included in the number equation, it is not in the $T_{\rm c}$ equation (\ref{gapETMA}). Indeed, the bare Green's function in Eq. (\ref{GAM0}) gives the effective Fermi surface size as, not ${\tilde k}_{{\rm F},\sigma}$, but 
\begin{equation}
{\tilde k}^0_{{\rm F},\sigma}=\sqrt{2m_\sigma\mu_\sigma},
\label{kfstar0}
\end{equation}
which remarkably depends on $\sigma={\rm L},{\rm H}$, as shown in Fig. \ref{fig4}(a). Thus, the $T_{\rm c}$-equation in the ETMA is affected by the mismatch of two Fermi surfaces (${\tilde k}_{{\rm F},{\rm L}}^0\ne {\tilde k}_{{\rm F},{\rm H}}^0$), leading to the suppression of the superfluid phase transition, as in the case of metallic superconductivity under an external magnetic field. To see this pair-breaking effect in a clear manner, it is convenient to rewrite Eq. (\ref{gapETMA}) in the form, 
\begin{equation}
1={U \over 2}
\sum_{\bm p}
{\displaystyle
\tanh{{\tilde \xi}_{{\bm p},{\rm L}}+h \over 2T}
+
\tanh{{\tilde \xi}_{{\bm p},{\rm H}}-h \over 2T}
\over
{\tilde \xi}_{{\bm p},{\rm L}}+{\tilde \xi}_{{\bm p},{\rm H}}
},
\label{ETMAgap2}
\end{equation}
where ${\tilde \xi}_{{\bm p},\sigma}=(m/m_\sigma)[p^2/(2m)-\mu]$, with $\mu=[\mu_{\rm L}+\mu_{\rm H}]/2$. Apart from the factor $m/m_{\sigma}$ in ${\tilde \xi}_{{\bm p},\sigma}$, Eq. (\ref{ETMAgap2}) has the same form as the $T_{\rm c}$-equation in a Fermi gas with an atomic mass $m$ and the Fermi chemical potential $\mu$, under an external magnetic field,
\begin{equation}
h={m_{\rm L}\mu_{\rm L}-m_{\rm H}\mu_{\rm H} \over m_{\rm L}+m_{\rm H}}.
\label{eq.h}
\end{equation}
When $(k_{\rm F}a_s)^{-1}=-0.5$, Fig. \ref{fig4}(b) shows that $T_{\rm c}$ in the ETMA disappears at $m_{\rm L}/m_{\rm H}\simeq 0.44$, at which one obtains ${\tilde k}_{{\rm F},{\rm L}}^0/k_{\rm F}=0.85$ and ${\tilde k}_{{\rm F},{\rm H}}^0/k_{\rm F}=0.61$. Substituting these into Eq. (\ref{eq.h}), we obtain $h=0.15T_{\rm F}$, which is comparable to the value of the superfluid phase transition temperature $T_{\rm c}=0.14T_{\rm F}$ at $m_{\rm L}/m_{\rm H}=1$. This clearly indicates that the absence of the superfluid phase when $m_{\rm L}/m_{\rm H}\le 0.44$ in Fig. \ref{fig4}(b) is due to the `magnetic field' $h$ in Eq. (\ref{eq.h}). However, since $h$ actually originates from the internal inconsistency of the ETMA, we conclude that the vanishing $T_{\rm c}$ seen in Fig. \ref{fig2}(a) is an artifact of this approximation. 
\par
We briefly note that the ETMA becomes consistent, when the dressed Green's function $G_\sigma$ in the number equation (\ref{number}) is replaced by the bare one $G_\sigma^0$ in Eq. (\ref{GAM0}). In this simple mean-field approximation, the number equation gives ${\tilde k}^0_{{\rm F}.{\rm L}}\simeq{\tilde k}^0_{{\rm F},{\rm H}}$, leading to $h\simeq 0$. Thus, we obtain a finite $T_{\rm c}$ for an arbitrary ratio $m_{\rm L}/m_{\rm H}$ of mass imbalance, as shown in Fig. \ref{fig4}(b) (although the magnitude of $T_{\rm c}$ is overestimated, because of the neglect of strong-coupling corrections).
\par
In the strong-coupling BEC regime, the system is well described by a Bose gas of $N/2$ tightly bound molecules\cite{Nozieres1985,Melo1993,Ohashi2002}, so that the difference between the ETMA and the SCTMA is not important, as far as we consider $T_{\rm c}$. Indeed, Fig. \ref{fig2} shows that the both approximations give almost the same $T_{\rm c}$ in the BEC regime. In this figure, we also compare our results with the BEC phase transition temperature $T_{\rm BEC}$ in an ideal gas of $N_{\rm B}=N/2$ hetero-molecules with the molecular mass $M=m_{\rm L}+m_{\rm H}$, given by
\begin{equation}
T_{\rm BEC}={2\pi \over M}\Bigl({N_{\rm B} \over \zeta(3/2)}\Bigr)^{2/3}.
\label{BEC}
\end{equation}
The good agreement of the SCTMA and ETMA results with $T_{\rm BEC}$ at $(k_{\rm F}a_s)^{-1}=2$ supports the validity of the molecular picture in this regime, even in the presence of mass imbalance.
\par
\begin{figure}
\begin{center}
\includegraphics[width=0.4\linewidth,keepaspectratio]{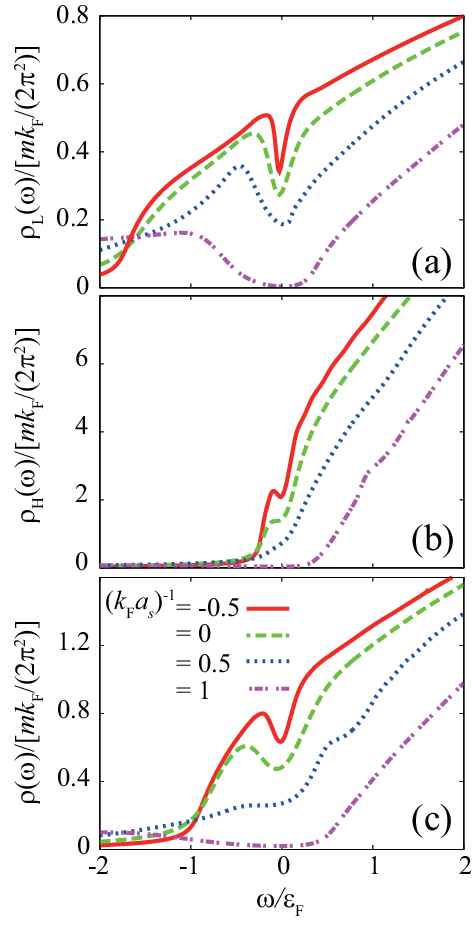}
\end{center}
\caption{(Color online) Calculated single-particle density of states $\rho_\sigma(\omega)$ in a mass-imbalanced Fermi gas at $T_{\rm c}$. We take $m_{\rm L}/m_{\rm H}=0.15$ (which corresponds to a $^6$Li-$^{40}$K Fermi mixture). (a) Light component. (b) Heavy component. For comparison, we also show the density of states $\rho(\omega)$ in the mass-balanced case in panel (c).
}
\label{fig5}
\end{figure}
\par
\section{Component-dependent pseudogap phenomena in a mass-imbalanced Fermi gas}\label{sec4}
\par
Figures \ref{fig5} shows the single-particle density of states $\rho_\sigma(\omega)$ at $T_{\rm c}$ in the case of a $^6$Li-$^{40}$K mixture ($m_{\rm L}/m_{\rm H}=0.15$). Panel (a) clearly shows that the density of states $\rho_{\rm L}(\omega)$ in the light component exhibits a dip structure when $(k_{\rm F}a_s)^{-1}=-0.5$, which becomes wider for a stronger pairing interaction. Since the superfluid order parameter vanishes at $T_{\rm c}$, this is just the pseudogap associated with pairing fluctuations. This result is qualitatively the same as the mass-balanced case shown in panel (c).
\par
Although both light atoms and heavy atoms equally contribute to pairing fluctuations (Note that a preformed Cooper pair always consists of a light atom and a heavy atom.), Fig. \ref{fig5}(b) shows that the pseudogap in the heavy component is not so clear as the case of light component. That is, a dip structure seen at $(k_{\rm F}a_s)^{-1}=-0.5$ no longer exists in the unitarity limit ($(k_{\rm F}a_s)^{-1}=0$), although a clear pseudogap is still seen in panel (a). In the BEC regime at $(k_{\rm F}a_s)^{-1}=1$, exactly speaking, there exists a wide pseudogap around $\omega=0$, which is, however, very shallow, so that it is almost invisible in this figure. This result is also quite different from the clear pseudogap structure seen in panel (a) at this interaction strength.
\par
\begin{figure}
\begin{center}
\includegraphics[width=0.9\linewidth,keepaspectratio]{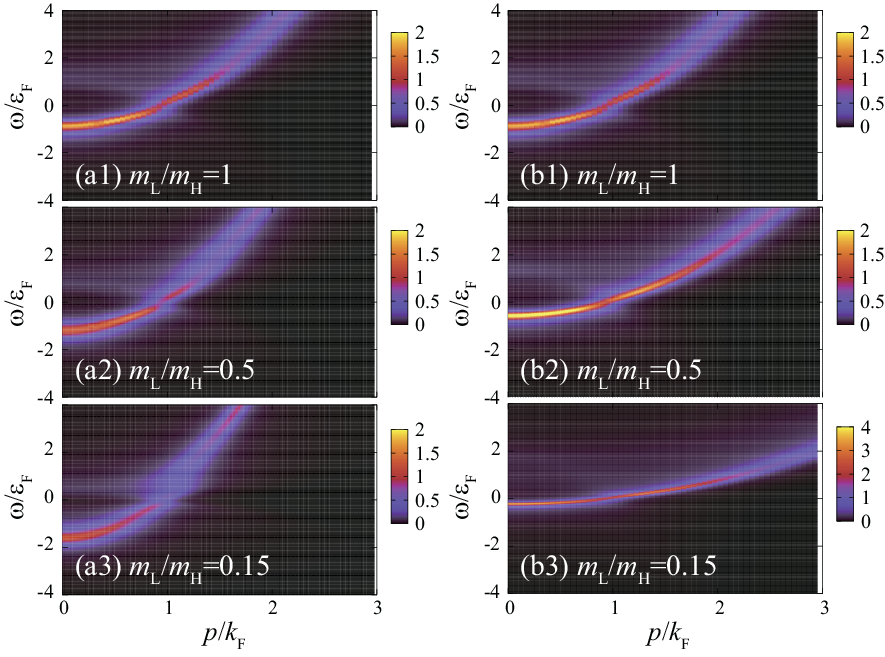}
\end{center}
\caption{(Color online) Calculated intensity of the spectral weight $A_\sigma({\bm p},\omega)$ in a unitary Fermi gas at $T=T_{\rm c}$, normalized by $\varepsilon_{\rm F}^{-1}$. The left and right panels show $A_{\rm L}(\bm p,\omega)$ and $A_{\rm H}(\bm p,\omega)$, respectively.}
\label{fig6}
\end{figure}
\par
Strong-coupling corrections to single-particle excitations can also be seen in the single-particle spectral weight $A_\sigma({\bm p},\omega)$ in Eq. (\ref{sp}). In the mass-balanced case, the pseudogap phenomenon appearing in the spectral weight may be understood as a particle-hole coupling effect by pairing fluctuations\cite{Tsuchiya2009}. Indeed, in Fig. \ref{fig6}(a1), besides a spectral peak line along the particle dispersion ($\xi_{\bm p}\sim p^2/(2m)-k_{\rm F}^2/(2m)$), we slightly see a broad peak line along the hole branch ($\xi^{\rm h}_{\bm p}\sim-[p^2/(2m)-k_{\rm F}^2/(2m)]$), which crosses the particle branch around $\omega=0$ to modify the particle dispersion. Since the density of states $\rho_\sigma(\omega)$ is obtained from the momentum summation of the spectral weight $A_\sigma({\bm p},\omega)$ for a given $\omega$, this modification around $\omega=0$ is directly related to the pseudogap structure in $\rho_\sigma(\omega)$ around $\omega=0$ (see Fig. \ref{fig5}). 
\par
In the light component, the same effect on the particle branch occurs in the presence of mass imbalance, as shown in Figs.\ref{fig6}(a2) and (a3). In particular, in the highly mass-imbalanced case (panel (a3)), the spectral peak of the particle branch is remarkably broadened around $\omega=0$ by the particle-hole coupling effect, leading to the suppression of the density of states $\rho_{\rm L}(\omega\sim 0)=\sum_{\bm p}A_{\rm L}({\bm p},\omega\sim 0)$, which gives the pseudogap structure in Fig. \ref{fig5}(a) at $m_{\rm L}/m_{\rm H}=0.15$.
\par
In the case of heavy atoms, on the other hand, the right panels in Fig. \ref{fig6} show that the modification of the particle branch around $\omega=0$ is less remarkable, compared with the case of light atoms. This result is consistent with the density of states shown in Fig. \ref{fig5}(b).
\par
\begin{figure}
\begin{center}
\includegraphics[width=0.5\linewidth,keepaspectratio]{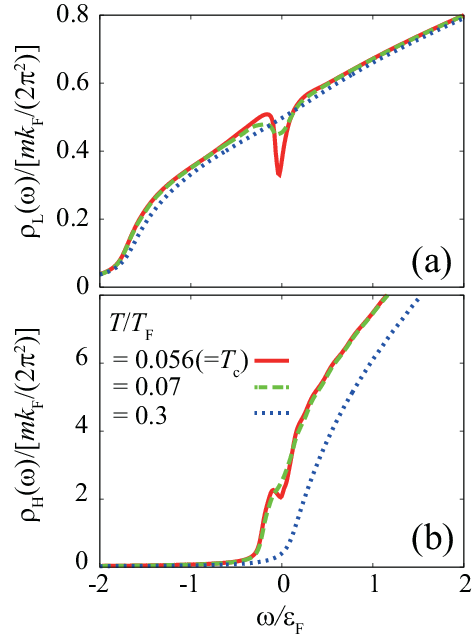}
\end{center}
\caption{(Color online) Single-particle density of states $\rho_\sigma(\omega)$ above $T_{\rm c}$. We take $m_{\rm L}/m_{\rm H}=0.15$ and $(k_{\rm F}a_s)^{-1}=-0.5$. (a) Light component. (b) Heavy component.}
\label{fig7}
\end{figure}
\begin{figure}[!ht]
\begin{center}
\includegraphics[width=0.95\linewidth,keepaspectratio]{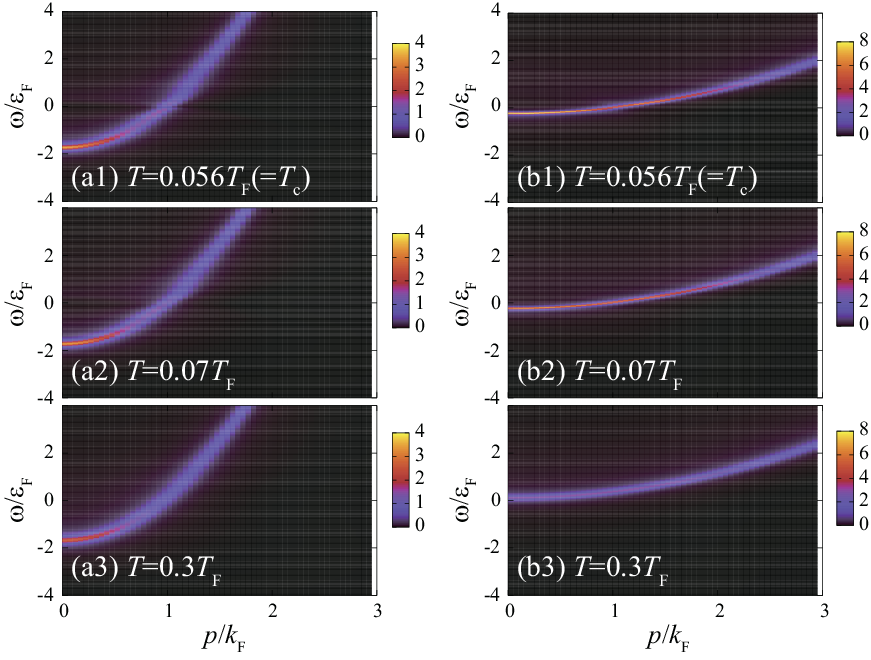}
\end{center}
\caption{(Color online) Single-particle spectral weight $A_\sigma(\bm p,\omega)$ above $T_{\rm c}$. (a1)-(a3) Light component. (b1)-(b3) Heavy component. We take  $m_{\rm L}/m_{\rm H}=0.15$, and $(k_{\rm F}a_s)^{-1}=-0.5$.}
\label{fig8}
\end{figure}

To understand the above component-dependent pseudogap phenomenon, we explain the following two keys. The first key is that the light atoms and heavy atoms have different Fermi temperatures as
\begin{equation}
T_{\rm F}^{\rm L}={k_{\rm F}^2 \over 2m_{\rm L}}>T_{\rm F}^{\rm H}={k_{\rm F}^2 \over 2m_{\rm H}}.
\label{eqTF}
\end{equation}
Since thermal effects in a Fermi gas are dominated by, not the temperature $T$ itself, but the {\it scaled} temperature $T/T_{\rm F}^\sigma$, heavy fermions always feel a higher scaled temperature than light fermions at a temperature $T$. Thus, the pseudogap in $\rho_{\rm H}(\omega)$ may be smeared out thermally, even when the pseudogap is still seen in $\rho_{\rm L}(\omega)$. In addition, since the difference of these scaled temperatures becomes larger for higher temperatures, the pseudogap in $\rho_{\rm H}(\omega)$ disappears at a lower temperature than in $\rho_{\rm L}(\omega)$. We explicitly confirm this in Fig. \ref{fig7} (density of states), as well as in Fig. \ref{fig8} (spectral weight). 
\par
\par
\par
\begin{figure}
\begin{center}
\includegraphics[width=0.5\linewidth,keepaspectratio]{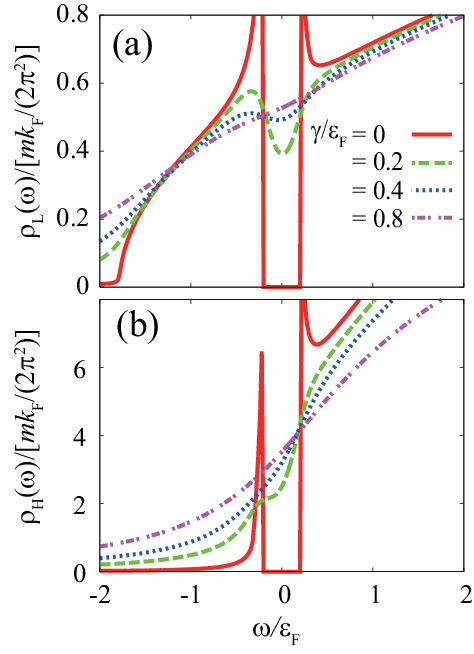}
\end{center}
\caption{(Color online) Density of states $\rho_\sigma(\omega)$, using the approximate Green's function ${\tilde G}^{\rm PG}_\sigma$ in Eq. (\ref{PGG3}). (a) Light component. (b) Heavy component. We take $m_{\rm L}/m_{\rm H}=0.15$, $\Delta_{\rm PG}/\varepsilon_{\rm F}=0.3$, and $\mu_\sigma=k_{\rm F}^2/(2m_\sigma)$. 
}
\label{fig9}
\end{figure}
\par
The second key to understand the component-dependent pseudogap phenomenon is the particle-hole coupling by pairing fluctuations. Noting that the particle-particle scattering matrix $\Gamma({\bm q}=0,i\nu_n=0)$ in Eq. (\ref{GAM}) diverges at $T_{\rm c}$, the self-energy $\Sigma_\sigma({\bm p},i\om _n)$ in Eq. (\ref{sctma}) can be approximated to, near $T_{\rm c}$\cite{Tsuchiya2009},
\begin{equation}
\Sigma_\sigma({\bm p},i\om _n)\simeq 
\Sigma_\sigma^{\rm Hartree}-G_{-\sig}(-{\bm p},-i\om_n)\Delta_{\rm pg}^2,
\label{sctma2}
\end{equation}
where $\Sigma_\sigma^{\rm Hartree}=-UT\sum_{{\bm p},i\omega_n}G_{-\sigma}({\bm p},i\omega_n)$ is the ordinary Hartree term, and $\Delta_{\rm pg}^2=-T\sum _{\bm q,\nu _n}[\Gam(\bm q,i\nu _n)+U]~(>0)$ is the so-called pseudogap parameter\cite{Chen2009}. When we simply treat the Green's function $G_{-\sigma}$ in Eq. (\ref{sctma2}) within the Hartree approximation ($[G_{-\sigma}(-\bm p, -i\omega_n)]^{-1}=-i\omega_n-\xi_{\bm p, -\sigma}-\Sigma_\sigma^{\rm Hartree}$), Eq. (\ref{sctmag}) is approximated to
\begin{eqnarray}
G^{\rm PG}_\sigma({\bm p},i\omega_n)=
{1 \over \displaystyle
i\omega_n-\xi_{{\bm p},\sigma}-
{\Delta_{\rm PG}^2 \over i\omega_n+\xi_{{\bm p},-\sigma}}
}.
\label{PGG}
\end{eqnarray}
(The unimportant Hartree term $\Sigma^{\rm Hartree}_\sigma$ has been absorbed into the Fermi chemical potential $\mu_\sigma$, for simplicity.) Equation (\ref{PGG}) indicates that the pseudogap parameter $\Delta_{\rm PG}$, which physically describes effects of pairing fluctuations, works as a coupling between a particle branch ($\omega=\xi_{{\bm p},\sigma}$) and a hole branch ($\omega=-\xi_{{\bm p},-\sigma}$). In addition, because Eq. (\ref{PGG}) can be written in the same form as the particle component of the BCS Green's function as
\begin{eqnarray}
G^{\rm PG}_\sigma({\bm p},i\omega_n)=
{i\omega_n+\xi_{{\bm p},-\sigma} \over
[i\omega_n-\xi_{{\bm p},\sigma}][i\omega_n+\xi_{{\bm p},-\sigma}]-\Delta_{\rm PG}^2},
\label{PGG2}
\end{eqnarray}
$\Delta_{\rm PG}$ is found to play a similar role to the BCS superfluid order parameter $\Delta$. Thus, the approximate Green's function in Eq. (\ref{PGG}) gives a BCS-like clear gap structure in both $\rho_{\rm L}(\omega)$ and $\rho_{\rm H}(\omega)$ (see $\rho_\sigma(\omega)$ at $\gamma=0$ in Fig. \ref{fig9}). 
\par
The component-dependent pseudogap phenomenon is then immediately obtained, when one phenomenologically includes finite widths of the peak lines in Figs.\ref{fig6} and \ref{fig8} as
\begin{eqnarray}
{\tilde G}^{\rm PG}_\sigma({\bm p},i\omega_n\to\omega+i\delta)=
{1 \over \displaystyle
\omega+i\gamma-\xi_{{\bm p},\sigma}-
{\Delta_{\rm PG}^2 \over \omega+i\gamma+\xi_{{\bm p},-\sigma}}
}.
\label{PGG3}
\end{eqnarray}
Here, the phenomenological damping rate $\gamma$ is assumed to take the same constant value between the two components, for simplicity. Using Eq. (\ref{PGG3}), one finds that the pseudogap in $\rho_{\rm H}(\omega)$ is more easily smeared out by the damping rate $\gamma$ than that in $\rho_{\rm L}(\omega)$, as shown in Fig. \ref{fig9}. When we simply consider the spectral weight $A_\sigma({\bm p},\omega=0)$ of the phenomenological Green's function $\tilde G_\sigma({\bm p},\omega)=[\omega+i\gamma-\xi_{{\bm p},\sigma}]^{-1}$, given by
\begin{equation}
A_\sigma({\bm p},\omega=0)=
{1 \over \pi}
{4m_\sigma^2\gamma \over [p^2-{\tilde k}^2_{{\rm F},\sigma}]^2
+4m_\sigma^2\gamma^2},
\label{PGG4}
\end{equation}
the density of states $\rho_\sigma(\omega=0)=\sum_{\bm p}A_\sigma({\bm p},\omega=0)$ is found to be dominated by the spectral weight in the momentum region, 
\begin{equation}
{\tilde k}^2_{{\rm F},\sigma}-2m_\sigma\gamma\lesssim p^2\lesssim{\tilde k}^2_{{\rm F},\sigma}+2m_\sigma\gamma.
\label{PGG5}
\end{equation}
This region is much wider for the heavy component than the light component, when $m_{\rm L}/m_{\rm H}\ll 1$. Thus, in the former, the modification of the particle dispersion around $p={\tilde k}_{{\rm F},{\rm H}}$ by the particle-hole coupling effect is easily hidden by the wider momentum summation, compared with the case of light component, leading to the different pseudogap phenomenon between the two.
\par
\begin{figure}
\begin{center}
\includegraphics[width=0.5\linewidth,keepaspectratio]{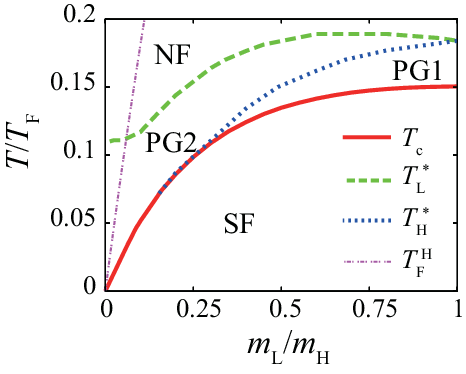}
\end{center}
\caption{(Color online) Pseudogap temperature $T_\sigma^*$ in a unitary Fermi gas, as a function of the ratio $m_{\rm L}/m_{\rm H}$ of mass imbalance. PG1: Pseudogap regime where a dip structure appears in both $\rho_{\rm L}(\omega)$ and $\rho_{\rm H}(\omega)$. PG2: Pseudogap region where the pseudogap is only seen in $\rho_{\rm L}(\omega)$. NF: Normal Fermi gas where the pseudogap phenomenon is absent. SF: Superfluid phase. $T_{\rm F}^{\rm H}$ is the Fermi temperature in the heavy component.
}
\label{fig10}
\end{figure}
\par
\section{Phase diagram of a mass-imbalanced Fermi gas}
\par
To determine the pseudogap region, we conveniently introduce the pseudogap temperature $T_{\sigma}^*$ as the temperature at which the pseudogap appears in the density of states $\rho_\sigma(\omega)$\cite{Tsuchiya2009}. As expected, Fig. \ref{fig10} shows that $T_{\rm L}^*>T_{\rm H}^*$. 
\par
This result naturally leads to the existence of two pseudogap regions. In the region $T_{\rm c}\le T\le T_{\rm H}^*$ (`PG1' in Fig. \ref{fig10}), the pseudogap appears in both the light component and heavy component. Besides this ordinary case, we also obtain the other pseudogap regime where the pseudogap only appears in the light component (`PG2' in Fig. \ref{fig10}). In Fig. \ref{fig10}, while PG1 and PG2 exist when $m_{\rm L}/m_{\rm H}>0.15$, PG2 only exists in the highly mass-imbalanced regime when $m_{\rm L}/m_{\rm H}\le 0.15$. Since PG2 is absent in the mass-balanced case, this pseudogap regime is characteristic of a mass-imbalanced Fermi gas. 
\par
\begin{figure}
\begin{center}
\includegraphics[width=0.5\linewidth,keepaspectratio]{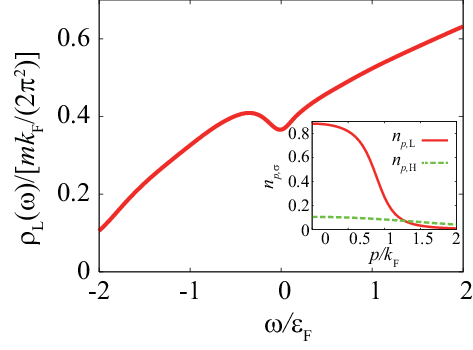}
\end{center}
\caption{(Color online) Pseudogapped density of states $\rho_{\rm L}(\omega)$, when heavy atoms are in the classical regime. We take $m_{\rm L}/m_{\rm H}=0.0125$, and $T/T_{\rm F}=0.05$, which satisfies $T_{\rm F}^{\rm H}=0.025T_{\rm F}<T<T_{\rm L}^*=0.11T_{\rm F}<T_{\rm F}^{\rm L}=1.98T_{\rm F}$. The inset shows the particle distribution $n_{{\bm p},\sigma}=\langle c^\dagger_{{\bm p},\sigma}c_{{\bm p},\sigma}\rangle$.
}
\label{fig11}
\end{figure}

Figure \ref{fig10} shows that the pseudogap temperature $T_{\rm L}^*$ in the light component becomes higher than $T_{\rm F}^{\rm H}$ in the case of $m_{\rm L}/m_{\rm H}\ll 1$. This means that the pseudogap phenomenon can still occur in the light component, even when the heavy component is in the classical regime ($T>T_{\rm F}^{\rm H}$). Indeed, as shown in Fig. \ref{fig11}, $\rho_{\rm L}(\omega)$ still exhibits the pseudogap phenomenon, when $T/T_{\rm F}=0.05$ (which satisfies $T_{\rm F}^{\rm H}=0.025T_{\rm F}<T<T_{\rm L}^*=0.11T_{\rm F}$)\cite{note3}. In this case, the inset of Fig. \ref{fig11} shows that the particle distribution $n_{{\bm p},{\rm H}}=\langle c_{{\bm p},{\rm H}}^\dagger c_{{\bm p},{\rm H}}\rangle$ of heavy atoms is very broad around $p=k_{\rm F}$, compared with $n_{{\bm p},{\rm L}}=\langle c_{{\bm p},{\rm L}}^\dagger c_{{\bm p},{\rm L}}\rangle$, reflecting that $T_{\rm F}^{\rm H}<T<T_{\rm F}^{\rm L}$. We briefly note that this result is in contrast to the case of superfluid phase transition, which occurs only when both the components are in the Fermi degenerate regime ($T_{\rm c}<T_{\rm F}^{\rm H}<T_{\rm F}^{\rm L}$).
\par
\par
\begin{figure}
\begin{center}
\includegraphics[width=0.5\linewidth,keepaspectratio]{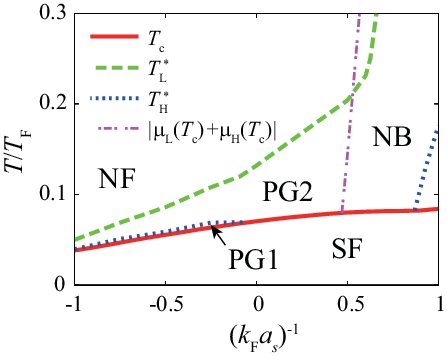}
\end{center}
\caption{(a) Phase diagram of a $^6$Li-$^{40}$K Fermi gas mixture ($m_{\rm L}/m_{\rm H}=0.15$). The meanings of PG1, PG2, SF, and NF, are the same as those in Fig. \ref{fig10}. In this figure, we also draw the line, $|\mu_{\rm L}(T_{\rm c})+\mu_{\rm H}(T_{\rm c})|$, in the BEC regime when $\mu_\sigma<0$, which physically represents the binding energy of a two-body bound molecules. As in the mass-balanced case, the right side of this line may be regarded as a molecular Bose gas in the normal state (NB), rather than a Fermi gas\cite{note2}.}
\label{fig12}
\end{figure}
\par
Figure \ref{fig12} shows the phase diagram of a $^6$Li-$^{40}$K mixture ($m_{\rm L}/m_{\rm H}=0.15$). As expected from Fig. \ref{fig10}, most of the pseudogap regime is dominated by PG2, where the pseudogap only appears in the light component. In the BEC limit, the molecular binding energy $E_{\rm bind}$ is given by
\begin{equation}
E_{\rm bind}=\mu_{\rm L}(T_{\rm c})+\mu_{\rm H}(T_{\rm c})=-{1 \over ma_s^2}.
\label{BECbind}
\end{equation}
Thus, in Fig. \ref{fig12}, the line $|\mu_{\rm L}(T_{\rm c})+\mu_{\rm H}(T_{\rm c})|$ (where $\mu_{\rm L}(T_{\rm c}) + \mu_{\rm H}(T_{\rm c})<0$) drawn in the BEC regime physically gives a characteristic temperature where two-body bound molecules start to appear. Thus, the right side of this line (NB) may be viewed as a normal Bose gas of two-body bound molecules, rather than a Fermi gas.
\par
\par
\begin{figure}
\begin{center}
\includegraphics[width=0.5\linewidth,keepaspectratio]{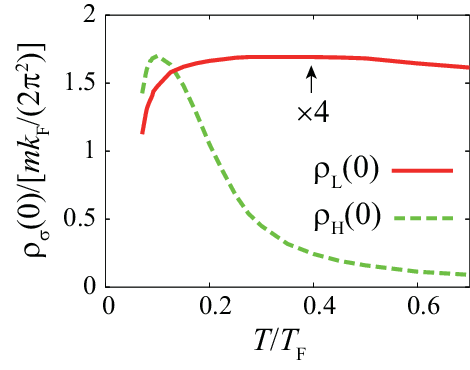}
\end{center}
\caption{(Color online) Calculated density of states $\rho_\sigma(\omega=0)$, as a function of temperature. We take $m_{\rm L}/m_{\rm H}=0.15$, and $(k_{\rm F}a_s)^{-1}=0$. $\rho_{\rm L}(\omega)$ is magnified four times.}
\label{fig13}
\end{figure}
\par
\begin{figure}
\begin{center}
\includegraphics[width=0.5\linewidth,keepaspectratio]{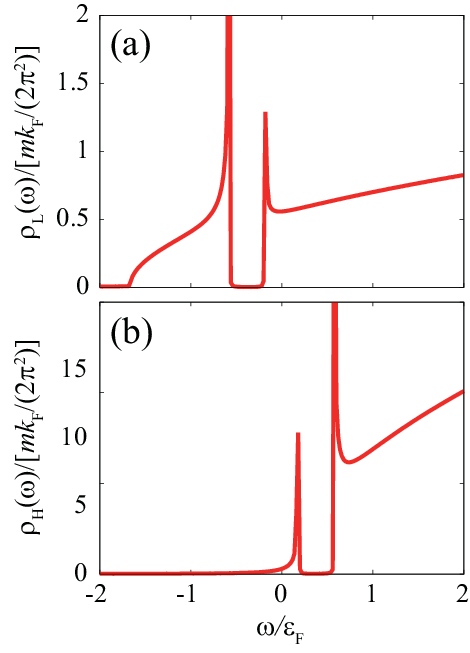}
\end{center}
\caption{(Color online) Superfluid density of states $\rho_\sigma(\omega)$ in the Sarma phase, calculated in the mean-field theory. (a) Light component. (b) Heavy component. We take $m_{\rm L}/m_{\rm H}=0.15$, $(k_{\rm F}a_s)^{-1}=0$, and $T=0.34T_{\rm F}~(=0.93T_{\rm c})$. }
\label{fig14}
\end{figure}
\par
Figure \ref{fig12} indicates that one should measure single-particle excitations in the light component, in order to observe the pseudogap phenomenon in a $^6$Li-$^{40}$K mixture. Since the achievement of the superfluid phase transition is a crucial issue in this system, this observation would be helpful to estimate to what extent we are approaching the superfluid instability. In addition, since the pseudogap is almost absent in the heavy component except for the very narrow temperature region (see Fig. \ref{fig12}), the appearance of a gap in single-particle excitation spectra in the heavy component would be a clear signature of the hetero superfluid state in this system. 
\par
We briefly note that, although Fig. \ref{fig12} indicates the absence of the pseudogap temperature $T_{\rm H}^*$ around the unitarity limit, it does not necessarily mean that the heavy component behaves as a simple normal Fermi gas there. Indeed, as shown in Fig. \ref{fig13}, in the unitarity limit, the density of states $\rho_{\rm H}(\omega)$ in the heavy component at $\omega=0$ is anomalously suppressed near $T_{\rm c}$ by strong pairing fluctuations, although a dip structure does not appear in $\rho_{\rm H}(\omega)$. Since the pseudogap is a crossover phenomenon without being accompanied by any phase transition, the definition of the pseudogap temperature somehow involves ambiguity. However, even when we define the pseudogap temperature $T_\sigma^*$ as the  temperature at which $\rho_\sigma(\omega=0)$ starts to be suppressed, we again obtain the relation $T_{\rm L}^*>T_{\rm H}^*$, as seen in Fig. \ref{fig13}.
\par
Before ending this section, we briefly comment on the Sarma phase\cite{Sarma1963}, which has been predicted in a highly mass-imbalanced Fermi gas\cite{Stoof1,Stoof2}. A characteristic property of this superfluid state is that the superfluid gap is not centered at $\omega=0$, as shown in Fig. \ref{fig14}. Since the pseudogap phenomenon in an ultracold Fermi gas is a precursor of the superfluid phase transition, one expects that the pairing fluctuations associated with the Sarma phase give a dip structure at $\omega\ne 0$. However, such a phenomenon is not seen in Fig. \ref{fig5}, where the pseudogap always appears around $\omega=0$. Thus, although this result does not necessarily exclude the Sarma phase in a mass-imbalanced Fermi gas, it seems difficult to confirm this possibility from the viewpoint of the pseudogap phenomenon.
\par
\par
\section{Summary}
\par
To summarize, we have discussed strong-coupling properties of an ultracold Fermi gas with different species with different masses. Extending our previous work using an extended $T$-matrix approximation (ETMA)\cite{Hanai2013} to include higher order pairing fluctuations within the framework of the self-consistent $T$-matrix approximation (SCTMA), we calculated the superfluid phase transition temperature $T_{\rm c}$ in the presence of mass imbalance in the whole BCS-BEC crossover region. We also calculated the single-particle density of states, as well as the single-particle spectral weight, to see how the presence of mass imbalance affects the pseudogap phenomenon.
\par
We showed that the superfluid phase transition always occurs even in the presence of mass imbalance. This result is quite different from our previous work within the ETMA, where the superfluid phase transition does not occur in the BCS regime when $m_{\rm L}/m_{\rm H}\ll 1$. We clarified that the ETMA result is an artifact, originating from the inconsistent treatment of the Fermi surface size between the $T_{\rm c}$-equation and the number equations $N_\sigma$ ($\sigma={\rm L},{\rm H}$). Our results in this paper predict that a $^6$Li-$^{40}$K mixture always exhibits the superfluid phase transition, irrespective of the interaction strength. Thus, the BCS-BEC crossover phenomenon is expected in this system, as in the cases of $^6$Li and $^{40}$K superfluid gases.
\par
We also showed that the pseudogap phenomena are very different between the light component and the heavy component, in spite that the both equally contribute to the formation of preformed Cooper pairs. In the presence of mass imbalance, the pseudogap structure in the density of states becomes obscure in the heavy component, compared with that in the light component. In the highly mass-imbalanced case ($m_{\rm L}/m_{\rm H}\ll 1$), the pseudogap no longer appears in the former. Since the pseudogap phenomenon always occurs in both the components in the mass-balanced case, this component-dependent pseudogap phenomenon is characteristic of a mass-imbalanced Fermi gas.
\par
The component-dependent pseudogap phenomenon also gives a higher pseudogap temperature $T_{\rm L}^*$ in the light component than the pseudogap temperature $T_{\rm H}^*$ in the heavy component, which naturally leads to two pseudogap regions. That is, while the both components exhibit the pseudogap phenomena when $T_{\rm c}\le T\le T_{\rm H}^*$, the pseudogapped density of states is only seen in the light component when $T_{\rm H}^*\le T\le T_{\rm L}^*$. In the highly mass-imbalance regime ($m_{\rm L}/m_{\rm H}\ll 1$), $T_{\rm H}^*$ no longer exists, so that light fermions only exhibit the pseudogap phenomenon there. We pointed out that that these component-dependent pseudogap phenomena originate from (1) different values of the Fermi temperatures between the two components, and (2) component-dependent particle-hole coupling effects by pairing fluctuations.
\par
For a $^6$Li-$^{40}$K mixture, our results predict that the pseudogap can be seen much more easily in the $^6$Li component, rather than in the $^{40}$K component, because this system is in the highly mass-imbalanced regime ($m_{\rm L}/m_{\rm H}=0.15\ll 1$). Since the pseudogap phenomenon is a precursor of the superfluid phase transition, the observation of this many-body phenomenon in the $^6$Li component would be helpful to assess to what extent the system is close to the superfluid instability. In addition, since the pseudogapped density of states is almost absent in the $^{40}$K component, the observation of a single-particle excitation gap in this component can be used as a signature of the hetero-superfluid phase in this system. 
\par
In this paper, we have examined a uniform Fermi gas, for simplicity. In this regard, we note that each component may feel different harmonic potential in a real trapped Fermi gas, leading to a local population (spin) imbalance\cite{Lin2006,Gezerlis2009}. In addition, the photoemission-type experiment developed by the JILA group\cite{Stewart2008,Gaebler2010}, which is a powerful technique to experimentally examine single-particle properties of an ultracold Fermi gas, has no spatial resolution, so that we need to treat an observed photoemission spectrum as a spatially averaged one in a trap. Thus, to deal with these realistic situations, the extension of our work to include effects of a harmonic trap is necessary. 
\par
Fermi superfluids with hetero-Cooper-pairs have been discussed in various fields, such as an exciton (polariton) gas in semiconductor physics, and color superconductivity in high-energy physics. Since the realization of a hetero pairing state seems difficult in metallic superconductivity, once the superfluid phase transition is achieved in a $^6$Li-$^{40}$K mixture, this superfluid state with a tunable pairing interaction would become a useful model system for the study of these Fermi condensates. Since the pseudogap phenomenon is deeply related to the superfluid phase transition, our results would contribute to the research toward the realization of a hetero Fermi superfluid using ultracold Fermi gases, especially a $^6$Li-$^{40}$K mixture.
\par
\begin{acknowledgements}
We thank H. Tajima for useful discussions. R.H. was supported by KLL PhD Program Research Grant, as well as Graduate School Doctoral Student Aid Program from Keio University. Y.O. was supported by Grant-in-Aid for Scientific research from MEXT in Japan (25105511, 25400418).
\end{acknowledgements}
\par


\begin{thebibliography}{99}
\bibitem{note} We briefly note there is actually no clear Fermi surface edge in the present case. The particle distribution $n_{{\bm p},\sigma}=\langle c_{{\bm p},\sigma}^\dagger c_{{\bm p},\sigma}\rangle$ is a smooth function with respect to the momentum $p=|{\bm p}|$, which gradually decreases to zero when $p\gesim{\tilde k}_{{\rm F},\sigma}$ in the Fermi degenerate regime.
\bibitem{note2} In Fig. \ref{fig12}, the pseudogap temperature $T_{\rm H}^*$ revives in the BEC regime. However, since the density of states $\rho_{\rm H}(\omega)$ is very small in the negative energy region (see, for example, Fig. \ref{fig5}(b)), the pseudogap is almost invisible there.
\bibitem{note3} Within our calculations, the pseudogap temperature $T_{\rm L}^*$ is always lower than $T_{\rm F}^{\rm L}$.
\end{thebibliography}
\end{document}